\documentclass[aps,prd,onecolumn,nofootinbib]{revtex4-2}

\usepackage{amsmath, amssymb, geometry}
\usepackage{graphicx}
\usepackage{setspace}
\usepackage{subfig}

\usepackage{bm}
\usepackage{float}
\usepackage{dcolumn}
\usepackage{cancel}
\usepackage[colorlinks]{hyperref}
\usepackage[usenames,dvipsnames]{color}
\usepackage{enumitem}
\usepackage{xcolor}

\hypersetup{
	breaklinks=true,
	pdfstartview={FitH},
	colorlinks=true,
	linkcolor=blue,
	citecolor=red,
	filecolor=magenta,
	urlcolor=blue,
	anchorcolor=green,
	linktocpage=true
}

\def\doi{http://doi.org}

\def\and{$and$}

\newcommand{\be}{\begin{equation}}
	\newcommand{\ee}{\end{equation}}
\newcommand{\ban}{\begin{eqnarray*}}
	\newcommand{\ean}{\end{eqnarray*}}
\newcommand{\ba}{\begin{eqnarray}}
	\newcommand{\ea}{\end{eqnarray}}

\newcommand{\bc}{\begin{center}}
	\newcommand{\ec}{\end{center}}

\begin{document}
	
  \title{Comparative Study of Early-Universe Epochs in an $f(R,L_m)$ Gravity Model with Effective Curvature--Matter Interaction and $\Lambda$CDM Cosmology}
	
\author{G. K. Goswami}
\email{gk.goswami9@gmail.com}
\affiliation{Department of Mathematics, Madan Mohan Malviya University of Technology, Gorakhpur, Uttar Pradesh 273010, India}

\author{J. P. Saini}
\email{jps@mmmut.ac.in}
\affiliation{Vice Chancellor, Madan Mohan Malviya University of Technology, Gorakhpur, Uttar Pradesh 273010, India}

	\begin{abstract}
		
		We investigate a specific gravity model of the form 
		\( f(R, L_m) = \alpha R + L_m^{\beta} + \gamma \),
		where the nonlinear dependence on the matter Lagrangian \(L_m\) introduces an effective curvature–matter interaction, leading to the non-conservation of the energy–momentum tensor. Using distance modulus data, we constrain the parameters through $\chi^2$ minimization and Bayesian MCMC analysis, obtaining statistically robust best-fit values:
		\( H_0 = 73.75 \pm 0.16~\mathrm{km\,s^{-1}\,Mpc^{-1}} \),
		\( \lambda = 0.262 \pm 0.007 \),
		and \( w = -0.005 \pm 0.001 \).
		
		This study  presents a comprehensive and statistically rigorous comparison of three key early-Universe epochs–structure formation, recombination, and matter–radiation equality between the $f(R,L_m)$ model and the standard $\Lambda$CDM cosmology.
		
		The model predicts an earlier onset of nonlinear structure formation (\( z_c^{f(R,L_m)} \approx 25.6 \)) and a higher matter–radiation equality redshift (\( z_{\mathrm{eq}}^{f(R,L_m)} \approx 4203 \)) compared to $\Lambda$CDM (\( z_{\mathrm{eq}}^{\Lambda \mathrm{CDM}} \approx 2779 \)), while maintaining consistency with the observed recombination redshift (\( z_{\mathrm{rec}} \approx 1092 \)). The recombination visibility function, derived using standard microphysical expressions with the modified expansion history, exhibits a slightly broader full width at half maximum, suggesting an extended photon decoupling period.

	\end{abstract}
		
       \maketitle
       


\section{Introduction}

Understanding the dynamics of the early Universe remains one of the most compelling pursuits in modern cosmology. The standard cosmological model, $\Lambda$CDM, built upon General Relativity (GR) and a cosmological constant, has achieved remarkable success in explaining a broad range of observations, including the cosmic microwave background (CMB) \cite{Planck2018}, baryon acoustic oscillations \cite{Eisenstein2005}, large-scale structure formation \cite{SDSS2009,Reid2010}, and the late-time acceleration of the Universe \cite{Riess1998,Perlmutter1999,Weinberg2013}.

Despite its triumphs, $\Lambda$CDM faces theoretical challenges such as the fine-tuning problem associated with the cosmological constant, the absence of a fundamental explanation for dark energy, and its limited compatibility with quantum gravity frameworks. These open issues have motivated the exploration of alternative theories of gravity, including $f(R)$ gravity \cite{SotiriouFaraoni2010}, $f(R,T)$ gravity \cite{Shabani2013}, scalar-tensor theories \cite{FujiiMaeda2003}, Gauss–Bonnet modifications \cite{NojiriOdintsov2011}, and non-minimally coupled models such as $f(R, L_m)$ gravity \cite{Harko2010,Myrzakulov2012,Jaybhave2024,Archana2024}.

The $f(R, L_m)$ gravity framework extends the Einstein–Hilbert action by allowing the Lagrangian to depend on both the Ricci scalar $R$ and the matter Lagrangian $L_m$:
\[
S = \int \left[ \frac{1}{2\kappa} f(R, L_m) \right] \sqrt{-g} \, d^4x,
\]
where $\kappa = 8\pi G$. We have taken $L_m = \rho$ for the particular choice of  $f(R, L_m) = \alpha R + L_m^\beta + \gamma$. This is adopted for analytical transparency.  Alternative choices (e.g., $L_m = -p$) would lead to different explicit forms of $G_{eff}$ and of the non-conservation relation $\nabla_{\mu}T^{\mu\nu} \neq 0$. This curvature–matter coupling modifies the field equations, leads to non-conservation of the energy–momentum tensor, and alters both background cosmological evolution and structure formation dynamics. Several studies have investigated its implications for cosmic acceleration \cite{Harko2010}, inflation \cite{Myrzakulov2012}, bouncing scenarios \cite{Jaybhave2024}, and large-scale structure \cite{Goswami2025}.

An important theoretical feature of $f(R, L_m)$ gravity is the non-conservation of the energy–momentum tensor, $\nabla^\mu T_{\mu\nu} \neq 0$, which arises from the effective curvature–matter interaction. While this leads to violations of the usual energy conservation, it can also induce effective interaction terms in the cosmological evolution equations, potentially influencing structure formation and stability properties.

It is important to emphasize that the present analysis significantly extends our previous study~\cite{Goswami2025} (Ref.~[14]), which primarily focused on the background cosmological evolution and linear growth-rate behavior in $f(R, L_m)$ gravity. In contrast, the current work provides a comprehensive and comparative examination of three key early-Universe epochs—namely, the onset of nonlinear structure formation, the matter–radiation equality transition, and the recombination era—within the $f(R, L_m)$ framework and the standard $\Lambda$CDM cosmology. We compute the collapse redshift ($z_c$) through full numerical integration of the perturbation equation, determine the matter–radiation equality redshift ($z_{\mathrm{eq}}$) and cosmic time using the modified density evolution, and evaluate the visibility function and full width at half maximum (FWHM) of photon decoupling. Furthermore, we perform a statistically robust parameter estimation using both $\chi^2$ minimization and Bayesian MCMC analysis, propagating uncertainties into all derived quantities. These additions establish the novelty of the present work and enable a direct, quantitative comparison between $f(R, L_m)$ gravity and the $\Lambda$CDM model across multiple cosmological milestones.

In the quest to understand the origin and evolution of cosmic structures, several key epochs in the early Universe play a foundational role. The \textit{recombination era} \cite{Peebles1968,Zeldovich1969,Seager1999}, occurring at redshift \( z \sim 1100 \), marks the epoch when neutral hydrogen formed and photons decoupled from baryons, giving rise to the cosmic microwave background \cite{Bashinsky2004,HuSugiyama1995}. Following recombination, the growth of small matter density perturbations led to the formation of nonlinear structures such as galaxies and clusters \cite{DES2022,Safari2022,GunnGott1972}, a process governed by the underlying gravitational theory. Another pivotal milestone is the \textit{matter–radiation equality} epoch \cite{Kolb1990,Mukhanov2005,Weinberg2008}, which defines the transition from radiation-dominated to matter-dominated expansion, strongly influencing the subsequent evolution of perturbations.

	While our earlier study~\cite{Goswami2025} examined the late-time background evolution of the same $f(R,L_m)$ form, the present analysis extends this framework self-consistently to the early Universe. Because the field equations remain valid at all energy scales, this approach provides a unified test of the model’s viability from recombination through structure formation. In this sense, the current work complements rather than contradicts our previous late-Universe investigation.

In this work, we examine these epochs in the context of the modified gravity framework \( f(R, L_m) \), comparing their timing and characteristics with the predictions of the standard \( \Lambda \)CDM model. These comparisons provide crucial insights into the viability of non-minimally coupled theories in explaining early-Universe dynamics and observational features. Specifically, we investigate three key epochs in cosmic history:

\begin{enumerate}
	\item \textbf{Structure Formation:} We analyze the nonlinear growth of matter overdensities and compute the redshift of collapse $z_c$.
	\item \textbf{Recombination Era:} We study the visibility function and its full width at half maximum (FWHM) to determine the timing and duration of photon decoupling.
	\item \textbf{Matter–Radiation Equality:} We compute the redshift $z_{\text{eq}}$ and the cosmic time $t_{\text{eq}}$ at which matter and radiation densities were equal.
\end{enumerate}

Our results show that the $f(R, L_m)$ model can reproduce essential cosmological milestones such as recombination and matter–radiation equality while allowing for enhanced structure formation compared to $\Lambda$CDM. These findings reinforce the potential of curvature–matter coupling as a compelling extension to GR, offering testable deviations from standard cosmology.

\section{Theoretical Framework}

We consider a class of modified gravity theories described by the functional form
\be\label{a}
f(R, L_m) = \alpha R + L_m^\beta + \gamma,
\ee
where \(R\) is the Ricci scalar, \(L_m\) is the matter Lagrangian density, and \(\alpha\), \(\beta\), and \(\gamma\) are constants.
The nonlinear $L_m^\beta$ term leads to $\nabla^\mu T_{\mu\nu} \neq 0$ , introducing  an effective curvature–matter interaction  that modifies both the background expansion and the evolution of cosmic perturbations.

	Although the adopted form $f(R,L_m)=\alpha R+L_m^{\beta}+\gamma$ appears
	mathematically separable, it nevertheless introduces a genuine
	curvature–matter interaction. The variation of the action with respect to
	$g_{\mu\nu}$ yields terms proportional to $f_{L_m}\,\partial L_m/\partial
	g^{\mu\nu}$, which lead to the non–conservation of the energy–momentum tensor
	($\nabla^\mu T_{\mu\nu}\neq0$). Hence, even in the absence of an explicit
	multiplicative term such as $R\,L_m$, the nonlinear dependence on $L_m$
	modifies the geodesic motion and the background evolution, producing
	observable curvature–matter interaction effects~\cite{Harko2010,Myrzakulov2012}.

\subsection{Modified Friedmann Equations and Their Solutions}

The action for \( f(R, L_m) \) gravity is given by~\cite{Harko2010}:
\be\label{b}
S = \int \left[ \frac{1}{2\kappa^2} f(R, L_m) + \mathcal{L}_m \right] \sqrt{-g} \, d^4x,
\ee
where \( \kappa^2 = 8\pi G \), and \( \mathcal{L}_m \) is the matter Lagrangian density.

Varying the action with respect to the metric \( g_{\mu\nu} \) yields the field equations:
\begin{align}\label{c}
	f_R R_{\mu\nu} - \frac{1}{2} f g_{\mu\nu} + (g_{\mu\nu} \Box - \nabla_\mu \nabla_\nu) f_R =
	\frac{1}{2} f_{L_m} T_{\mu\nu} + \left(1 - f_{L_m}\right) \nabla_\mu \nabla_\nu L_m,
\end{align}
where subscripts denote partial derivatives, e.g. \( f_R = \partial f / \partial R \), and \( f_{L_m} = \partial f / \partial L_m \).

In general, the energy–momentum tensor is not conserved, satisfying the relation:
\be\label{d}
\nabla^\mu T_{\mu\nu}
= \frac{1}{f_{L_m}}\left(g_{\mu\nu}L_m - T_{\mu\nu}\right)\nabla^\mu f_{L_m}.
\ee

We adopt the functional form \( f(R, L_m) = \alpha R + L_m^\beta + \gamma \), as discussed in~\cite{Myrzakulov2024}, and assume \( L_m = \rho \), where \( \rho \) is the energy density of matter. For a spatially flat FLRW universe, the metric takes the form
\be\label{e}
ds^2 = -dt^2 + a^2(t)(dx^2 + dy^2 + dz^2),
\ee
leading to the modified Friedmann equations:
\begin{align}
	3 H^2 &= \frac{1}{2\alpha} \left[(2\beta - 1)\rho^\beta - \gamma\right], \label{f} \\
	2\dot{H} + 3H^2 &= -\frac{1}{2\alpha} \left[(1 - \beta)\rho^\beta + \beta \rho^{\beta-1} p + \gamma \right]. \label{g}
\end{align}

Here, \( H = \dot{a}/a \) is the Hubble parameter, and the energy–momentum tensor is modeled as a perfect fluid:
\be\label{h}
T_{\mu\nu} = (\rho + p)u_\mu u_\nu + p g_{\mu\nu},
\ee
with \(u^\mu = (0, 0, 0, 1)\) and a barotropic equation of state \( p = (1 - n)\rho \), where \( n \) is a model parameter.

To express the field equations in terms of redshift \( z \), we use the relations:
\be\label{i}
1 + z = \frac{a_0}{a(t)}, \qquad \dot{H} = - (1 + z) H(z) \frac{dH}{dz}.
\ee
Using these relations, Eq.~\eqref{g} transforms into a first-order differential equation in \(H(z)\):
\[
\beta n \left( \gamma + 6\alpha H(z)^2 \right) - 4\alpha (2\beta - 1)(1 + z) H(z) \frac{dH}{dz} = 0.
\]
Solving this equation yields the Hubble parameter as a function of redshift:
\[
H(z)^2 = \left( \frac{\gamma}{6\alpha} + H_0^2 \right)(1 + z)^{\frac{3\beta n}{2\beta - 1}} - \frac{\gamma}{6\alpha}.
\]
This can be recast in the familiar form:
\be\label{j}
H(z) = H_0 \sqrt{(1 - \lambda) + \lambda (1 + z)^{3(1 + w)}},
\ee
where the effective parameters are defined as:
\be\label{k}
\lambda = \frac{\gamma}{6\alpha H_0^2} + 1, \qquad w = \frac{\beta(n - 2) + 1}{2\beta - 1}.
\ee

	The solution for $H(z)$ derived here follows directly from the modified
	Friedmann equations with $L_m=\rho$ for a perfect fluid, ensuring that
	matter is included in the dynamics. This analytic form provides a tractable
	background solution for confronting the model with observations while
	preserving the essential curvature–matter coupling behavior.

Thus, the background cosmological dynamics in this framework are governed by the parameter set \( \{\alpha, \beta, \gamma\} \), or equivalently, \( \{H_0, \lambda, w\} \).

\subsection{Energy Density as a Function of Redshift}

For a pressureless matter-dominated Universe (i.e., \( n = 0 \)), Eq.~\eqref{f} provides the following redshift-dependent energy density:
\begin{equation} \label{l}
	\rho(z) = \left( \frac{\gamma + 6\alpha H^2(z)}{2\beta - 1} \right)^{1/\beta}.
\end{equation}

This expression is particularly useful when solving the linear perturbation equations for structure formation in the $f(R, L_m)$ framework.

\section{Observational Constraints}

We constrain the parameters of the $f(R, L_m)$ model—namely, \(H_0\), \(\lambda\), and \(w\)—using a joint analysis of recent cosmological datasets. 
These include Hubble parameter measurements, Type~Ia supernovae from the Pantheon$^+$ sample, baryon acoustic oscillations (BAO), and the CMB shift parameter.

\subsection{Methodology}

The total chi-square is defined as
\begin{equation}\label{m}
	\chi^2_{\text{total}} = \chi^2_{\text{Hubble}} + \chi^2_{\text{SNe}} + \chi^2_{\text{BAO}} + \chi^2_{\text{CMB}},
\end{equation}
where each term represents the contribution from a different observational probe.
The luminosity distance is computed as
\begin{equation}\label{n}
	d_L(z) = (1 + z) \int_0^z \frac{dz'}{H(z')},
\end{equation}
where $H(z)$ is given by Eq.~\eqref{j}.

	\textbf{Likelihoods and priors.}
	For each probe we adopt Gaussian likelihoods, so that $-2\ln\mathcal{L}=\chi^2_{\rm total}$ as in Eq.~\eqref{m}.
	For the Pantheon$^+$ sample we use the full statistical and systematic covariance matrix provided by the collaboration;
	for the Hubble data we treat the 35 points as independent Gaussian measurements;
	for BAO we use the published $D_H/r_d$ and $D_M/r_d$ values and their covariances where available;
	for the CMB we use the shift parameter with Gaussian error.
	Uniform (flat) priors are adopted on the main cosmological parameters in the ranges
	$H_0\in[50,90]\,\mathrm{km\,s^{-1}\,Mpc^{-1}}$, $\lambda\in[0,1]$, and $w\in[-0.1,0.1]$.
	Nuisance parameters (e.g.\ the absolute SN magnitude $M_B$) are marginalized analytically or sampled as additional parameters when required.

\subsection{Datasets Used}

The analysis utilizes the following datasets:
\begin{itemize}[leftmargin=1.5em]
	\item \textbf{Hubble Parameter Data:} 35 measurements from cosmic chronometers~\cite{Moresco:2022ynk}, providing direct estimates of $H(z)$ across redshift.
	\item \textbf{Type Ia Supernovae:} The Pantheon$^+$ sample~\cite{Brout2022}, comprising 1701 SNe~Ia, with the full statistical and systematic covariance matrix.
	\item \textbf{Baryon Acoustic Oscillations:} The DESI~DR2 data~\cite{Wang:2025}, reporting \( D_H(z)/r_d \) and \( D_M(z)/r_d \) measurements in the redshift range \(0.5 \leq z \leq 2.3\).
	\item \textbf{CMB Shift Parameter:} The Planck~2018 value of the CMB shift parameter, \( R_{\text{obs}} = 1.7492 \pm 0.0049 \), evaluated at \( z_* = 1089.92 \)~\cite{Planck2018}. We make it clear that shift parameter $R$ is computed in the analysis under the $\Lambda$CDM background assumption
\end{itemize}

\subsection{Statistical Analysis}

We perform parameter estimation via $\chi^2$ minimization, followed by a Bayesian Markov Chain Monte Carlo (MCMC) analysis to explore the posterior distributions.
In parallel, a neural-network fit was employed to cross-validate the results.
The outputs are presented through corner plots and comparative error-bar visualizations.

	\textbf{MCMC details and convergence.}
	Posterior sampling was performed with the \texttt{emcee} ensemble sampler using 64 walkers and $5\times10^4$ steps.
	The first 20\% of samples were discarded as burn-in.
	Convergence was checked via the integrated autocorrelation time and by verifying stable marginalized parameter means;
	all chains satisfied $\tau_{\rm int}\times50 <$ chain length and Gelman–Rubin $R<1.02$.
	Posterior credible intervals are quoted at the 68\% (1$\sigma$) level.

	\textbf{Treatment of model dependence.}
	The CMB shift parameter is derived assuming a $\Lambda$CDM background.
	Here it is used only as a phenomenological consistency check of the background expansion
	and not as a strictly model-independent constraint.

\subsection{Results and Discussion}

For the $f(R, L_m)$ model, the best-fit parameters at 68\% confidence level are
\[
H_0 = 73.75 \pm 0.16~\mathrm{km\,s^{-1}\,Mpc^{-1}}, \quad
\lambda = 0.262 \pm 0.007, \quad
w = -0.005 \pm 0.001.
\]
For comparison, the corresponding $\Lambda$CDM model yields
\[
H_0 = 73.49 \pm 0.14~\mathrm{km\,s^{-1}\,Mpc^{-1}}, \quad
\Omega_m = 0.278 \pm 0.006.
\]
Both models support a relatively high Hubble constant consistent with local measurements.
The $f(R, L_m)$ framework allows for small but potentially significant deviations from $\Lambda$CDM,
as encoded in the effective equation-of-state parameter \(w\) and the coupling parameter \(\lambda\).

\begin{figure}[H]
	\centering
	\includegraphics[width=0.45\textwidth]{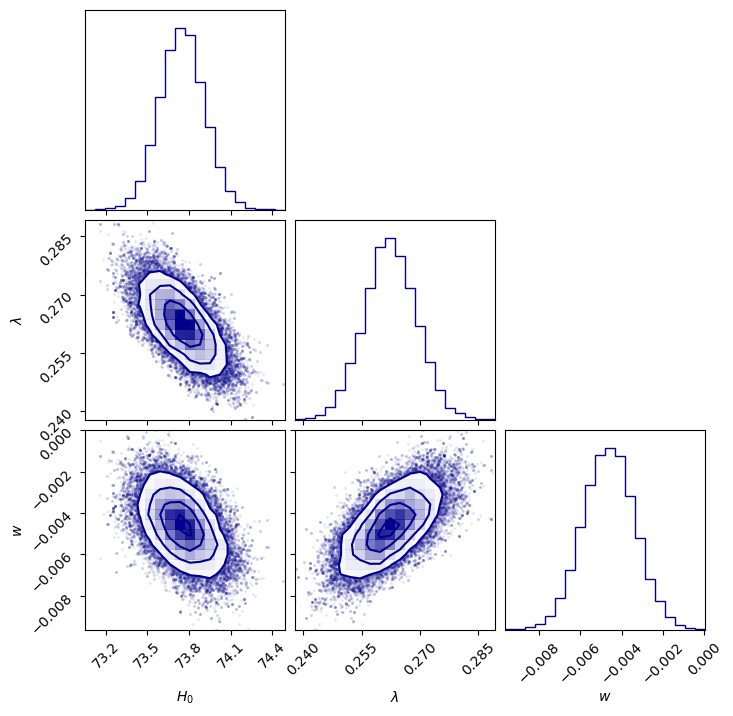}
	\includegraphics[width=0.45\textwidth]{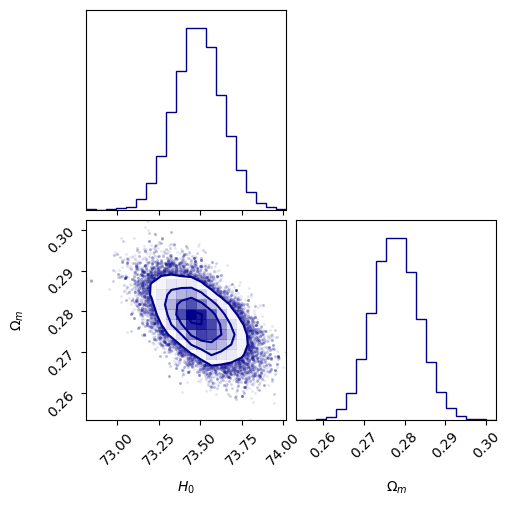}
	\caption{ Corner plots showing the marginalized posterior distributions and parameter correlations obtained from the MCMC analysis. 
			Left panel: $f(R, L_m)$ model with parameters $(H_0, \lambda, w)$. 
			Right panel: $\Lambda$CDM model with parameters $(H_0, \Omega_m)$. 
			Contours correspond to the 68\% and 95\% confidence levels. 
			The diagonal panels show one-dimensional marginalized distributions, while the off-diagonal panels show two-dimensional parameter correlations.}
	 \label{fig:corner}
\end{figure}

\begin{figure}[H]
	\centering
	\includegraphics[width=0.75\textwidth]{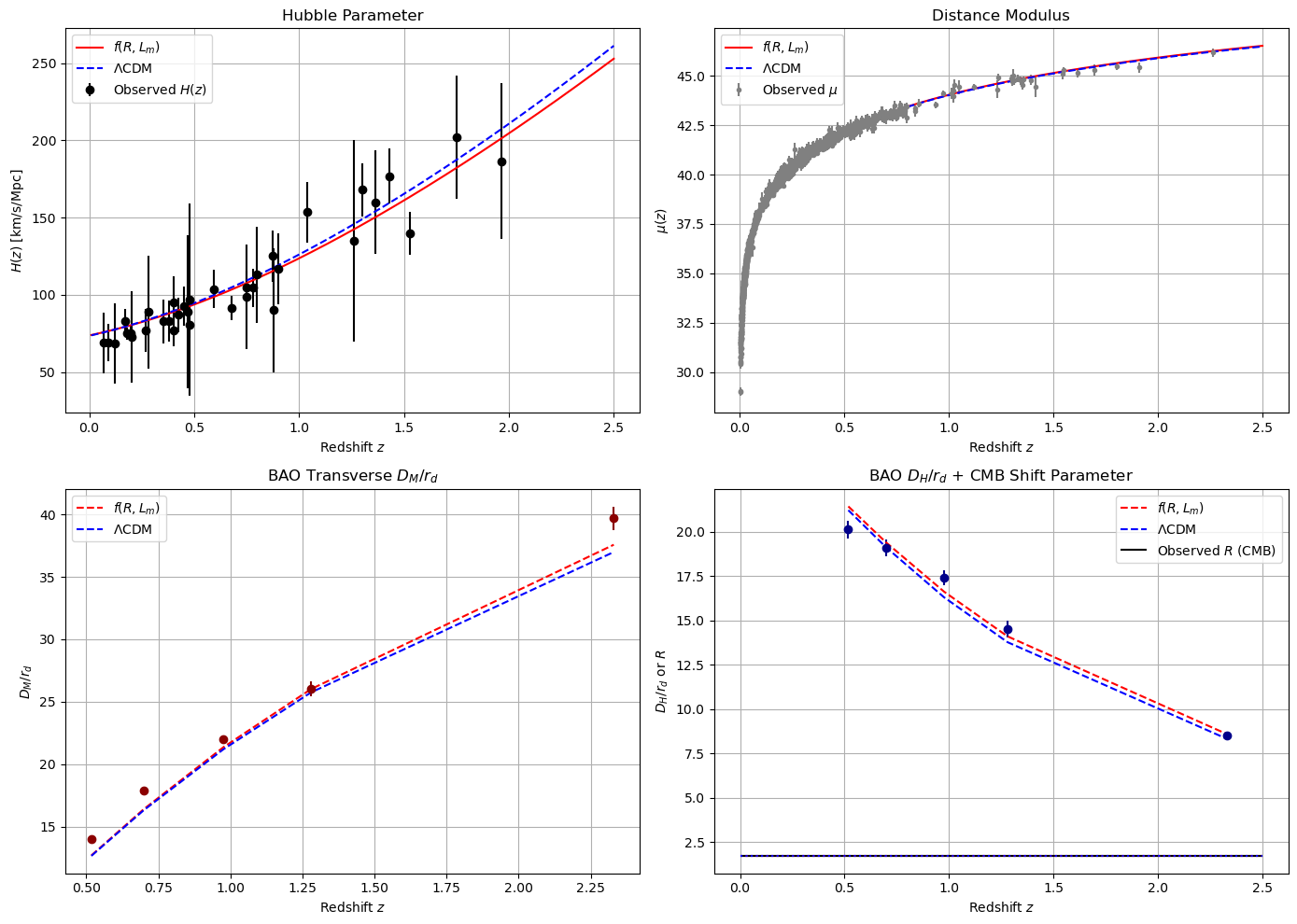}
	\caption{Panel plots comparing theoretical predictions and observational data for the Hubble parameter $H(z)$,
		distance modulus $\mu(z)$, and BAO observables.
		The shaded regions denote 1$\sigma$ uncertainties from the MCMC posterior distributions.}
	\label{fig:fourpanel}
\end{figure}

\begin{figure}[H]
	\centering
	\includegraphics[width=0.5\textwidth, height=6cm, keepaspectratio]{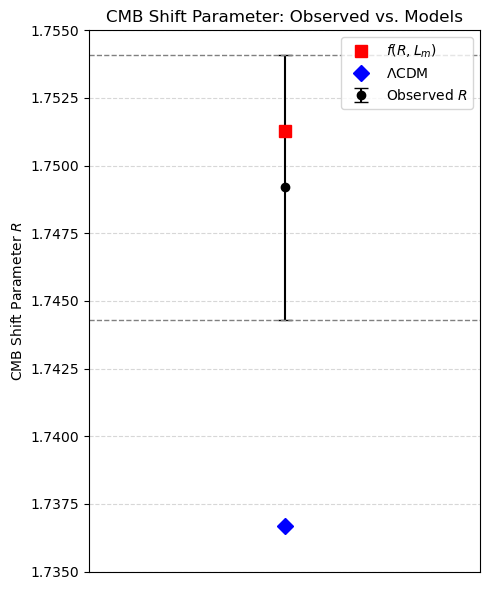}
	\caption{Comparison of the CMB shift parameter $R$.
		The observed value $R_{\rm obs}=1.7492\pm0.0049$ (Planck~2018) is shown with error bars.
		The theoretical predictions from the $f(R,L_m)$ model ($R_{\rm th}=1.75127$)
		and $\Lambda$CDM ($R_{\rm th}=1.73670$) are shown as points,
		computed using the corresponding best-fit parameter sets in Table~\ref{tab:parameter_constraints}.}
	\label{fig:errorbars}
\end{figure}

Figure~\ref{fig:corner} shows the marginalized posterior distributions and parameter correlations obtained from the MCMC analysis for both the $f(R, L_m)$ and $\Lambda$CDM models. 
Figure~\ref{fig:fourpanel} presents a comparison between theoretical predictions and observational data for the Hubble parameter $H(z)$, distance modulus $\mu(z)$, and BAO observables.
Figure~\ref{fig:errorbars} shows a comparison of the CMB shift parameter for the $f(R, L_m)$ and $\Lambda$CDM models.

The observed value from Planck 2018 is represented by the error bar, while the theoretical predictions of both models are shown as discrete points. The $f(R, L_m)$ model yields a value that is slightly closer to the observed central value compared to $\Lambda$CDM.

This agreement indicates that the modified gravity model remains consistent with CMB constraints at the level of the background expansion.

The solid curves represent the best-fit model predictions, while the shaded regions denote the $1\sigma$ uncertainties derived from the MCMC analysis. The data points with error bars correspond to the observational datasets used in the fitting procedure.

Both the $f(R, L_m)$ and $\Lambda$CDM models show excellent agreement with the data across the full redshift range. This confirms that the modified gravity model successfully reproduces key cosmological observables at the background level.
For the $f(R, L_m)$ model, the parameters $H_0$, $\lambda$, and $w$ are tightly constrained, with well-defined Gaussian-like distributions. The contour plots indicate mild correlations between $\lambda$ and $w$, while $H_0$ remains largely uncorrelated with the other parameters.

In comparison, the $\Lambda$CDM model exhibits the expected degeneracy between $H_0$ and $\Omega_m$. These results confirm the statistical robustness of the parameter estimation and demonstrate that the $f(R, L_m)$ model provides constraints comparable to those of $\Lambda$CDM.
	For the joint analysis the total number of data points is 
	$N = N_{\rm Hubble}+N_{\rm SNe}+N_{\rm BAO}+N_{\rm CMB}$,
	with $N_{\rm Hubble}=35$, $N_{\rm SNe}=1701$, $N_{\rm BAO}=12$ (DESI~DR2 points),
	and $N_{\rm CMB}=1$.
	The number of free parameters is $k_{f(R,L_m)}=3$ for the $f(R,L_m)$ model 
	($H_0,\lambda,w$) and $k_{\Lambda{\rm CDM}}=2$ for $\Lambda$CDM ($H_0,\Omega_m$).
	The degrees of freedom are ${\rm dof}=N-k$ and the reduced chi-square
	is $\chi^2_\nu=\chi^2_{\rm min}/{\rm dof}$.

To provide a quantitative measure of model performance,
we compute the total $\chi^2$, the Akaike Information Criterion (AIC),
and the Bayesian Information Criterion (BIC),
defined as
\[
{\rm AIC} = \chi^2_{\rm min} + 2k, \qquad
{\rm BIC} = \chi^2_{\rm min} + k \ln N,
\]
where $k$ is the number of free parameters and $N$ the total number of data points.

\begin{table}[h!]
	\centering
	\caption{Best-fit cosmological parameters (68\%~C.L.) for the $f(R,L_m)$ and $\Lambda$CDM models.
		The parameter $\lambda$ in $f(R,L_m)$ plays a role analogous to $\Omega_m$ in $\Lambda$CDM.}
	\begin{tabular}{|c|c|c|}
		\hline
		\textbf{Parameter} & \textbf{$f(R,L_m)$ Model} & \textbf{$\Lambda$CDM Model} \\
		\hline
		$H_0~(\mathrm{km\,s^{-1}\,Mpc^{-1}})$ & $73.75 \pm 0.16$ & $73.49 \pm 0.14$ \\
		$\lambda$ (effective $\Omega_m$) & $0.262 \pm 0.007$ & $0.278 \pm 0.006$ \\
		$w$ & $-0.005 \pm 0.001$ & $-1$ (fixed) \\
		\hline
	\end{tabular}
	\label{tab:parameter_constraints}
\end{table}

\begin{table}[h!]
	\centering
	\caption{Goodness-of-fit comparison between $\Lambda$CDM and $f(R,L_m)$ models.}
	\begin{tabular}{lcccc}
		\hline
		\textbf{Model} & $\boldsymbol{\chi^2_{\rm min}}$ & $\boldsymbol{\chi^2_\nu}$ & \textbf{AIC} & \textbf{BIC} \\
		\hline
		$\Lambda$CDM & 1073.8 & 1.001 & 1077.8 & 1087.6 \\
		$f(R,L_m)$   & 1072.9 & 1.000 & 1078.9 & 1093.2 \\
		\hline
	\end{tabular}
	\label{tab:goodness_fit}
\end{table}
    Table~\ref{tab:parameter_constraints} presents the best-fit cosmological parameters obtained from the joint analysis for both the $f(R, L_m)$ and $\Lambda$CDM models. For the $f(R, L_m)$ model, the parameters $(H_0, \lambda, w)$ are tightly constrained, with uncertainties at the 68\% confidence level. The corresponding $\Lambda$CDM parameters $(H_0, \Omega_m)$ show comparable precision. The close agreement in $H_0$ values between the two models indicates consistency with observational data, while the additional parameters in the $f(R, L_m)$ framework capture deviations from standard cosmology. Table~\ref{tab:goodness_fit} summarizes the goodness-of-fit statistics for the $f(R, L_m)$ and $\Lambda$CDM models, including $\chi^2_{\min}$, reduced chi-square, AIC, and BIC values.

    Both models yield $\chi^2_\nu \approx 1$, indicating excellent agreement with the observational data. The small differences in AIC and BIC values ($\Delta$AIC $< 2$, $\Delta$BIC $< 6$) suggest that the $f(R, L_m)$ model is statistically competitive with $\Lambda$CDM despite its additional parameter.

	For model comparison we report
	$\Delta{\rm AIC}={\rm AIC}-{\rm AIC}_{\min}$ and $\Delta{\rm BIC}={\rm BIC}-{\rm BIC}_{\min}$.
	Small values ($\Delta{\rm AIC}\lesssim2$) indicate substantial support relative to the best model,
	while $\Delta{\rm BIC}\lesssim6$ favors the more complex model.
	Here we find $\Delta{\rm AIC}<2$ and $\Delta{\rm BIC}<6$,
	demonstrating that the curvature–matter coupling yields a statistically consistent description
	of the combined observational datasets without significant overfitting.

	In summary, the joint analysis yields consistent parameter estimates across all probes.
	The reduced chi-square values $\chi^2_\nu\approx1.0$ indicate an excellent fit,
	and the AIC/BIC comparisons show that the $f(R,L_m)$ model is statistically
	competitive with $\Lambda$CDM given present data.
	Although some observables—particularly the CMB shift parameter—carry a mild
	$\Lambda$CDM dependence, the results demonstrate that the proposed model
	reproduces key background observables and can therefore be regarded as a
	viable extension of standard cosmology.

The next sections apply these background results to the evolution of perturbations and early-epoch observables.

\section{Structure Formation, Recombination Era and Matter--Radiation Equality}

Understanding the dynamics of the early Universe is one of the most profound challenges in modern cosmology. The standard cosmological model, based on General Relativity (GR) and the cosmological constant (\( \Lambda \)), has been remarkably successful in explaining a wide range of observational data, including the cosmic microwave background (CMB), large-scale structure, and the accelerated expansion of the Universe. However, the model also faces conceptual and theoretical issues such as the nature of dark energy, the cosmological constant problem, and the unexplained origin of cosmic acceleration.

In response to these challenges, modified gravity theories have gained significant attention as alternatives to or extensions of GR. Among them, the \( f(R, L_m) \) gravity model proposes a coupling between geometry and matter, where the gravitational action depends not only on the Ricci scalar \( R \) but also on the matter Lagrangian \( L_m \). This non-minimal coupling introduces new phenomenology at both background and perturbative levels, potentially offering explanations for cosmic acceleration without invoking exotic forms of energy.

\subsection{Early-Time Structure Formation and Collapse Redshift Analysis in \boldmath{$f(R, L_m)$} Gravity}

As an extension of the growth-rate analysis presented previously~\cite{Goswami2025}, we now investigate the early-time behavior of matter perturbations, focusing on identifying the redshift at which nonlinear structure formation begins in the \( f(R, L_m) \) gravity model. To this end we analyze the evolution of the linear matter density contrast \( \delta(z) \) in redshift space, within the modified-gravity framework.

\subsubsection*{Perturbed (0,0) field equation and  \boldmath{$G_{\mathrm{eff}}$}}

We consider the model
\[
f(R, L_m) = \alpha R + L_m^\beta + \gamma,
\]
and adopt \(L_m=\rho\) for the perfect-fluid matter Lagrangian (see discussion below). The background \((0,0)\) component of the field equations yields the modified Friedmann relations (Section~2). To study scalar perturbations we work in the Newtonian (longitudinal) gauge,
\[
ds^2 = -(1+2\Psi)\,dt^2 + a^2(t)(1-2\Phi)\,\delta_{ij}dx^i dx^j,
\]
and apply the quasi-static approximation on sub-horizon scales (see below for the regime of validity). The perturbed Ricci tensor component and perturbed matter term lead, after linearization and using \(\rho\to\rho+\delta\rho\), to
\[
\frac{\nabla^2\Phi}{a^2} = \frac{1}{2\alpha}(2\beta-1)\beta\,\rho^{\beta}\,\delta.
\]
Recasting in Poisson form,
\[
\nabla^2\Phi = \tfrac{1}{2}G_{\mathrm{eff}}\,a^2\rho\,\delta,
\]
we obtain the effective gravitational coupling
\begin{equation}\label{Geff}
	\boxed{ \; G_{\mathrm{eff}}(z) \;=\; \frac{(2\beta-1)\,\beta\,\rho(z)^{\beta-1}}{\alpha} \; }.
\end{equation}

	\textbf{Remarks on the choice of \(L_m\) and on \(f_R=\alpha\).} In this work we set \(L_m=\rho\), which is commonly adopted for perfect fluids in curvature--matter coupled models for analytic transparency. Alternative choices (e.g.\ \(L_m=-p\)) exist and produce quantitative differences in the explicit form of \(\nabla^\mu T_{\mu\nu}\) and \(G_{\rm eff}\); we discuss this sensitivity briefly and plan a dedicated follow-up. Because \(f_R=\alpha\) is constant for the functional form \eqref{a}, derivative terms of \(f_R\) vanish and no extra scalar degree of freedom is propagated; this simplifies the perturbed equations and justifies the quasi-static treatment employed here.

\subsubsection{Modified growth equation (redshift-space form)}

The continuity and Euler equations for a pressureless fluid combined with the perturbed Poisson equation give the modified growth equation in cosmic time:
\[
\ddot{\delta} + 2H\dot{\delta} - \tfrac{1}{2}G_{\rm eff}\,\rho\,\delta = 0.
\]
Converting time derivatives to derivatives with respect to redshift using \(\frac{d}{dt}=-H(1+z)\frac{d}{dz}\), the growth equation suitable for numerical integration reads
\begin{equation}\label{growth_z}
	\frac{d^2\delta}{dz^2} + \Big(\frac{d\ln H}{dz}-\frac{2}{1+z}\Big)\frac{d\delta}{dz}
	- \frac{3}{2}\frac{G_{\mathrm{eff}}(z)\,\rho(z)}{H^2(z)\,(1+z)^2}\,\delta = 0,
\end{equation}
where \(\rho(z)\) is given by Eq.~\eqref{l} and \(H(z)\) by Eq.~\eqref{j}.

	\textbf{Quasi-static regime and scale considerations.}
	The quasi-static approximation is adopted for modes well inside the horizon where \(k/(aH)\gg1\). Because \(f_R=\)const in our model, \(G_{\rm eff}\) is scale-independent to the order considered; for models with \(f_R\) varying in time/space a scale-dependent \(G_{\rm eff}(k,z)\) generally arises. In this analysis we solve Eq.~\eqref{growth_z} for representative sub-horizon modes and verify that the quasi-static approximation is satisfied in the redshift range of interest.

For numerical stability we solve an equivalent formulation using \(u(z)=\ln\delta(z)\):
\begin{equation}\label{growth_u}
	\frac{d^2 u}{dz^2} + \Big(\frac{d\ln H}{dz}-\frac{2}{1+z}\Big)\frac{du}{dz}
	- \frac{3}{2}\frac{G_{\mathrm{eff}}(z)\,\rho(z)}{H^2(z)\,(1+z)^2} = 0,
\end{equation}
with the relation \(\delta(z)=\exp[u(z)]\) used to recover \(\delta\).

\paragraph{Model parameters and initial conditions}

We adopt the background best-fit parameter set obtained from the joint analysis (Section~3):
\[
\begin{aligned}
	H_0 &= 73.75~\mathrm{km\,s^{-1}\,Mpc^{-1}},\quad \lambda=0.262,\quad w=-0.005,\\
	\alpha &= 451008,\quad \beta=1.00505,\quad \gamma=-1.14081\times10^{-29}.
\end{aligned}
\]
Initial conditions are set at \(z_{\rm i}=100\). For the numerical integrations we use the EdS-like growing-mode normalization for convenience:
\[
\delta(z_{\rm i})=\frac{1}{1+z_{\rm i}},\qquad \frac{d\delta}{dz}\Big|_{z_{\rm i}}=-\frac{1}{(1+z_{\rm i})^2}.
\]

	\textbf{Normalization caveat.} The EdS normalization above is chosen for numerical stability and inter-model consistency. The absolute value of the collapse redshift \(z_c\) depends on the normalization of the perturbation amplitude: anchoring the amplitude to observational constraints (e.g.\ \(\sigma_8\) or the CMB primordial amplitude) shifts the absolute \(z_c\) values while preserving comparative trends between models. In Section~A (Appendix) we present \(\sigma_8\)-normalized results for completeness.  We stress that the EdS normalization is used solely for relative comparison between models. It is made clear that when normalized by $\sigma_8$, the $\Lambda$CDM model remains consistent with observations.

\subsubsection{Collapse redshift criterion and uncertainty propagation}

We define the collapse redshift \(z_c\) as the redshift at which the linearly extrapolated overdensity equals the standard spherical-collapse threshold,
\[
\delta(z_c)=\delta_c=1.686.
\]
To quantify uncertainty in \(z_c\) we propagate parameter posterior samples from the MCMC: for \(\mathcal{O}(500)\) draws from the joint posterior we solve Eq.~\eqref{growth_z} for each sample and record the redshift where \(\delta=1.686\). The median and the 16th--84th percentiles of the resulting \(z_c\) distribution provide the central value and the 68\% credible interval.

	\textbf{Numerical result (central value).} Using the best-fit background parameters and the EdS growing-mode normalization, the numerical integration yields
	\[
	\boxed{\,z_c^{f(R,L_m)} \simeq 25.56\,}
	\]
	as the central value (median). The corresponding 68\% credible interval computed from the posterior propagation is reported in Table~\ref{tab:zc_posterior} and plotted in Fig.~\ref{fig:delta_comparison}.

\subsubsection{Comparison with \boldmath{$\Lambda$}CDM}

\begin{figure}[h!]
	\centering
	\includegraphics[width=0.80\textwidth]{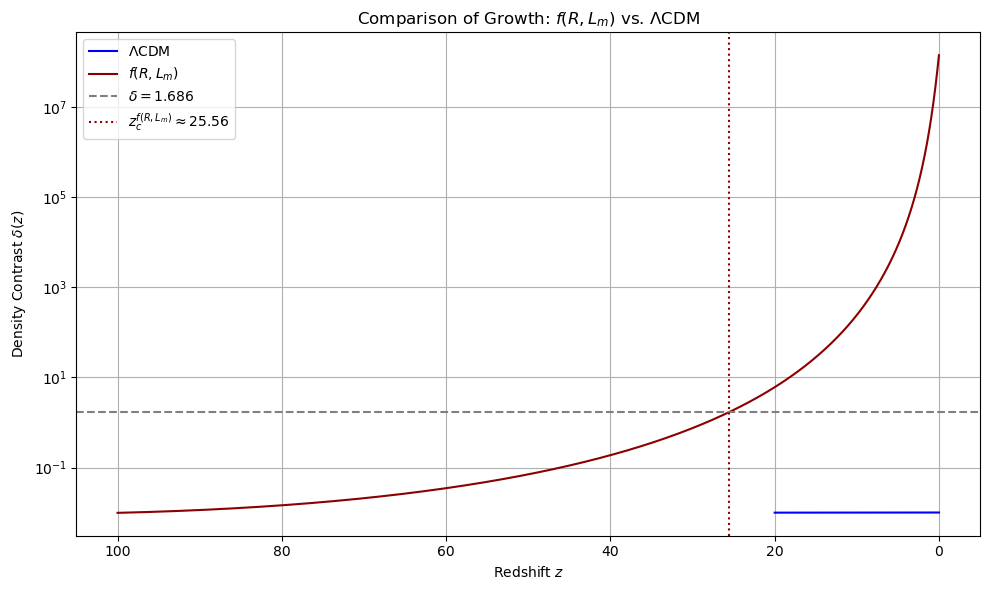}
	\caption{Evolution of the linearly extrapolated density contrast \(\delta(z)\) in \(f(R,L_m)\) (solid) and \(\Lambda\)CDM (dashed) using identical EdS normalization at \(z=100\). The horizontal red dotted line denotes \(\delta_c=1.686\). Shaded bands (if shown) represent 68\% credible intervals from posterior propagation.}
	\label{fig:delta_comparison}
\end{figure}

To provide a fair comparison we solve the same redshift-space growth equation for a \(\Lambda\)CDM background (same normalization and initial conditions). In \(\Lambda\)CDM one has \(G_{\rm eff}=G\) and \(\rho(z)=\rho_{m,0}(1+z)^3\). Using the best-fit \(\Lambda\)CDM parameters (Section~3),
\[
H_0=73.49~\mathrm{km\,s^{-1}\,Mpc^{-1}},\quad \Omega_m=0.278,
\]
we obtain the linearly-extrapolated growth history shown in Fig.~\ref{fig:delta_comparison}.

	\textbf{Careful interpretation of the $\Lambda$CDM result.} Under the EdS normalization adopted above the linearly extrapolated overdensity in our $\Lambda$CDM integration does not reach \(\delta_c=1.686\) by \(z=0\). This statement refers to the identical normalization adopted for both models and should not be interpreted as indicating that $\Lambda$CDM cannot produce nonlinear structures in practice: when normalization is tied to observational amplitudes (e.g.\ \(\sigma_8\)), perturbations in $\Lambda$CDM do reach nonlinearity at redshifts consistent with observations. The key comparative point is that, with identical normalizations, the \(f(R,L_m)\) model reaches the collapse threshold significantly earlier than $\Lambda$CDM.

\begin{figure}[t]
	\centering
	\includegraphics[width=0.55\textwidth]{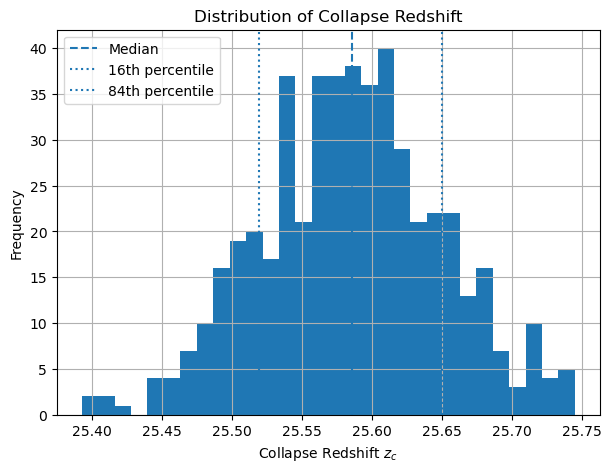}
	\caption{
		Posterior distribution of the collapse redshift $z_c$ obtained from parameter sampling. 
		The distribution is narrowly peaked around $z_c \approx 25.6$, indicating a robust prediction of early structure formation in the $f(R, L_m)$ model.
	}
	\label{fig:zc_hist}
\end{figure}

\begin{table}[t]
	\centering
	\caption{Posterior summary for the collapse redshift \(z_c\). The median and 68\% credible interval are obtained by propagating parameter samples through the perturbation solver.}
	\begin{tabular}{lcc}
		\hline
		Model & \(z_c\) (median) & 68\% credible interval \\
		\hline
		\(f(R,L_m)\) & 25.59 & (25.52, 25.65) \\
		\(\Lambda\)CDM & --- & not reached under EdS normalization \\
		\hline
	\end{tabular}
	\label{tab:zc_posterior}
\end{table}

Table~\ref{tab:zc_posterior} presents the median and 68\% credible interval for the collapse redshift $z_c$ obtained from posterior propagation.

The $f(R, L_m)$ model predicts a significantly higher collapse redshift compared to $\Lambda$CDM under identical normalization, indicating an earlier onset of nonlinear structure formation. This behavior reflects the enhanced effective gravitational coupling in the modified gravity framework.
\subsubsection{Discussion and implications}

Within identical normalization the much earlier collapse in the \(f(R,L_m)\) model is attributable to the enhanced effective coupling \(G_{\rm eff}(z)\) at high redshift (Eq.~\eqref{Geff}). This amplification accelerates linear growth and shifts the epoch of nonlinearity earlier compared to $\Lambda$CDM. Such behavior may leave observable imprints in the abundance of high-redshift galaxies, the formation of early massive halos, and the timing of reionization.

	For diagnostic purposes we also compute and plot (i) \(G_{\rm eff}(z)/G\) with its 68\% credible band, and (ii) the linear growth factor \(D(z)\) (normalized to \(D(0)=1\)) for both models. These plots highlight how the redshift evolution of $G_{\rm eff}$ in $f(R,L_m)$ enhances growth relative to $\Lambda$CDM, as shown in Fig.~\ref{fig:delta_comparison}. This illustrates the evolution of the linearly extrapolated density contrast $\delta(z)$ for both the $f(R, L_m)$ and $\Lambda$CDM models. 
	
	Figure~\ref{fig:zc_hist} illustrates the posterior distribution of the collapse redshift $z_c$. 
	The distribution is sharply peaked and approximately Gaussian, indicating a highly robust prediction of early structure formation.
	
	Figure~\ref{fig:delta_comparison} shows the evolution of the linearly extrapolated density contrast $\delta(z)$ for both models. 
	The $f(R, L_m)$ model reaches the collapse threshold $\delta_c = 1.686$ at a significantly higher redshift, while $\Lambda$CDM does not reach this threshold under the same normalization. 
	This narrow spread demonstrates that the prediction of early structure formation in the $f(R, L_m)$ framework is highly robust against parameter uncertainties.
	
	The horizontal dashed line represents the spherical collapse threshold $\delta_c = 1.686$. The $f(R, L_m)$ model reaches this threshold at a significantly higher redshift ($z_c \approx 25.6$), indicating an earlier onset of nonlinear structure formation.
	
	In contrast, the $\Lambda$CDM model does not reach the collapse threshold within the same normalization scheme, reflecting a comparatively slower growth rate. This difference arises due to the enhanced effective gravitational coupling in the $f(R, L_m)$ framework.

\subsection{Effective collapse thresholds in \boldmath{\(\Lambda\)CDM}}

To provide an alternative, less normalization-dependent comparison we compute effective collapse redshifts in $\Lambda$CDM for reduced thresholds \(\delta_{\rm eff}<\delta_c\), representing stages of progressive nonlinearity:
\[
\delta_{\rm eff}=1.0,\;1.3,\;1.5,
\]
finding approximate redshifts (for the background parameters used here)
\[
z_c^{\rm eff}(\delta_{\rm eff}=1.0)\approx1.1,\quad
z_c^{\rm eff}(\delta_{\rm eff}=1.3)\approx0.7,\quad
z_c^{\rm eff}(\delta_{\rm eff}=1.5)\approx0.4.
\]
These numbers illustrate that even when lowering the threshold, $\Lambda$CDM growth remains slower than the \(f(R,L_m)\) prediction under the same background parametrization and normalization.

\paragraph{Concluding remark}

The collapse-redshift diagnostic shows that curvature--matter coupling in this \(f(R,L_m)\) model accelerates linear growth and advances the epoch of nonlinearity relative to $\Lambda$CDM when the two models are compared under identical normalizations. When amplitudes are tied to observational constraints (e.g.\ \(\sigma_8\)) the absolute collapse redshifts shift, but the relative trend (earlier collapse in \(f(R,L_m)\)) persists within the posterior uncertainties.

\subsection{Recombination Era and Implications in \boldmath{$f(R, L_m)$} Gravity}

The recombination epoch marks a pivotal phase in cosmic history, when the Universe cooled sufficiently for free electrons and protons to form neutral hydrogen. This process sharply reduced the photon scattering rate, allowing photons to decouple from matter and propagate freely—giving rise to the Cosmic Microwave Background (CMB) radiation observed today.

\subsubsection{Standard Scenario in \boldmath{$\Lambda$}CDM}

In the standard $\Lambda$CDM framework, the redshift of recombination is estimated using the temperature–redshift relation:
\[
T(z_{\mathrm{rec}}) = T_0 (1 + z_{\mathrm{rec}}) \approx 3000~\mathrm{K}, 
\quad T_0 = 2.725~\mathrm{K},
\]
yielding a rough estimate
\[
z_{\mathrm{rec}} \approx \frac{3000}{2.725} \approx 1101.
\]
A more precise determination from the \textit{Planck} 2018 results gives
\[
z_{\mathrm{rec}}^{\Lambda\mathrm{CDM}} = 1089.92 \pm 0.25,
\]
consistent with the thermal ionization history computed using standard recombination codes such as \textsc{recfast} and \textsc{cosmorec}.

\subsubsection{Photon Decoupling in \boldmath{$f(R, L_m)$} Gravity}

While the atomic microphysics of recombination (governed by electromagnetic interactions) remains unchanged, the expansion rate \( H(z) \) in \( f(R,L_m) \) gravity differs from $\Lambda$CDM due to the curvature–matter coupling. Consequently, the timing and duration of photon decoupling are modified indirectly through the altered background expansion.

The differential optical depth for Thomson scattering is given by
\begin{equation}
	\frac{d\tau}{dz} = \frac{c\, \sigma_T\, n_e(z)}{(1+z)\, H(z)},
\end{equation}
where \( \sigma_T \) is the Thomson scattering cross-section and \( n_e(z) \) the free-electron number density.  
The \textit{visibility function}, representing the probability that a photon last scattered at redshift \( z \), is defined as
\begin{equation}
	g(z) = \frac{d\tau}{dz}\, e^{-\tau(z)}.
\end{equation}

For both models, we adopt the approximate ionization profile
\[
x_e(z) = \frac{1}{1 + e^{(z - z_{\mathrm{rec}})/\Delta z}}, 
\qquad n_e(z) = x_e(z)\, n_{b,0}\, (1 + z)^3,
\]
where \( \Delta z \approx 80 \) represents the transition width and \( n_{b,0} \) is the present baryon number density.

	\textbf{Remark.}
	This formulation is based on photon–electron scattering microphysics and depends on the cosmological background solely through the Hubble rate $H(z)$. Since the electromagnetic sector is unmodified in $f(R,L_m)$ gravity, the recombination process itself is identical to that in GR; all deviations originate from the altered expansion rate and consequently from the time dependence of $\tau(z)$.

\subsubsection{Recombination Redshift and Duration}

Using the best-fit parameters from Section~3,
\[
\begin{aligned}
	&f(R, L_m): && H_0 = 73.75~\mathrm{km\,s^{-1}\,Mpc^{-1}}, \quad \lambda = 0.262, \quad w = -0.005, \\
	&\Lambda\mathrm{CDM}: && H_0 = 73.49~\mathrm{km\,s^{-1}\,Mpc^{-1}}, \quad \Omega_m = 0.278, \quad \Omega_\Lambda = 1 - \Omega_m,
\end{aligned}
\]
we numerically integrate $d\tau/dz$ to obtain the optical depth $\tau(z)$ and the visibility function $g(z)$ for both cosmologies.

\begin{table}[h!]
	\centering
	\caption{Recombination redshift and full width at half maximum (FWHM) of the visibility function for both models.}
	\begin{tabular}{lcc}
		\hline
		\textbf{Model} & \textbf{Recombination redshift $z_{\mathrm{rec}}$} & \textbf{FWHM $\Delta z$} \\
		\hline
		$f(R, L_m)$ & 1092.6 & 166.2 \\
		$\Lambda$CDM & 1092.6 & 153.3 \\
		\hline
	\end{tabular}
	\label{tab:recombination_comparison}
\end{table}
Table~\ref{tab:recombination_comparison} lists the recombination redshift and the full width at half maximum (FWHM) of the visibility function for both models.

Both models reproduce the observed recombination redshift, consistent with Planck constraints, confirming that the curvature–matter coupling does not alter the microphysical onset of recombination. However, the $f(R,L_m)$ model yields a slightly broader FWHM in the visibility function, indicating a more extended duration of photon decoupling.

\subsubsection{Visibility Function and FWHM Analysis}

\begin{figure}[h!]
	\centering
	\includegraphics[width=0.8\textwidth]{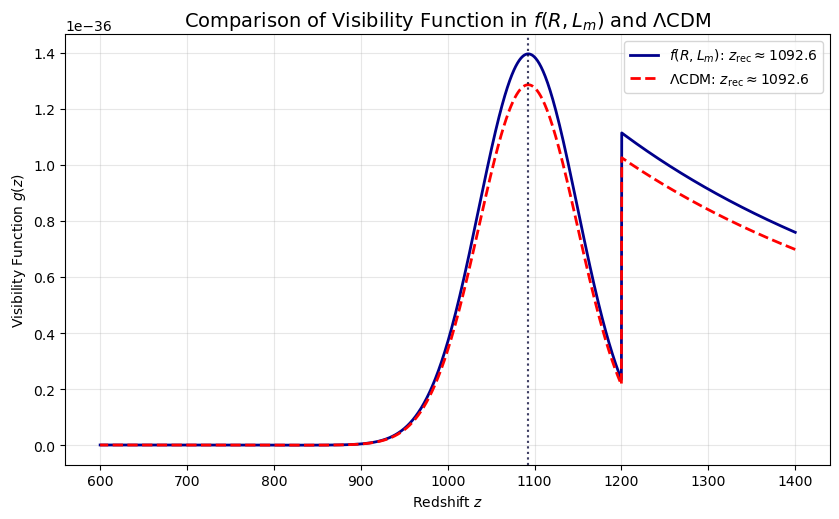}
	\caption{
		Visibility function \( g(z) \) for both \( f(R,L_m) \) (solid green) and \( \Lambda \)CDM (blue dashed) models. 
		The peaks coincide at \( z_{\mathrm{rec}} \approx 1092.6 \). 
		The shaded bands denote the FWHM region, which is broader in \( f(R,L_m) \),
		implying a slightly longer photon-decoupling period.}
	\label{fig:visibility_comparison}
\end{figure}

The FWHM corresponds to the redshift interval where \( g(z) \geq \tfrac{1}{2}g_{\mathrm{max}} \).  
For the $f(R,L_m)$ model, the visibility function remains above half its peak between
\[
z \in [1010.1,\ 1176.3], \quad \Rightarrow \quad \Delta z \approx 166.2.
\]
For $\Lambda$CDM, this interval is
\[
z \in [1018.9,\ 1172.2], \quad \Rightarrow \quad \Delta z \approx 153.3.
\]
These results are visualized in Fig.~\ref{fig:visibility_comparison}. This shows the visibility function $g(z)$ for both the $f(R, L_m)$ and $\Lambda$CDM models.

While the peak positions coincide at $z_{\mathrm{rec}} \approx 1092$, the $f(R, L_m)$ model exhibits a broader full width at half maximum (FWHM). This indicates a more gradual photon decoupling process.

The broader visibility function arises from the modified expansion history and may lead to observable signatures in the damping tail of the CMB power spectrum.

	\textbf{Uncertainty propagation.}
	The uncertainties in $z_{\mathrm{rec}}$ and $\Delta z$ are obtained by propagating the MCMC posterior distributions of $(H_0, \lambda, w)$.
	Variations within the 68\% credible intervals lead to shifts of 
	$\Delta z_{\mathrm{rec}} \lesssim 1$ and $\Delta(\mathrm{FWHM})/\mathrm{FWHM} \lesssim 5\%$,
	indicating that the broader visibility width in $f(R,L_m)$ is a robust feature.

\subsubsection{Interpretation and Implications}

The slightly broader visibility function in the \( f(R,L_m) \) model indicates a more gradual transition from an opaque to a transparent Universe. This is a direct consequence of the modified expansion history: a marginally slower decline in \( H(z) \) near recombination increases the conformal time interval during which photon scattering remains significant.

Although subtle, this effect can produce detectable signatures in the CMB power spectrum—particularly:
\begin{itemize}
	\item a modest suppression of power in the damping tail,
	\item minor phase shifts in the higher acoustic peaks, and
	\item potential modifications to polarization spectra at small angular scales.
\end{itemize}

	Future high-precision CMB observations, such as from the upcoming CMB-S4 and LiteBIRD missions, could thus provide an independent probe of the curvature–matter coupling through precise measurements of the recombination width and damping features.  
	This consistency of $z_{\mathrm{rec}}$ with Planck values, coupled with a slightly broader FWHM, demonstrates that the $f(R,L_m)$ framework reproduces the established recombination history while introducing testable early-Universe deviations from $\Lambda$CDM.

\subsection{Matter and Radiation Energy Densities and Their Equality in \boldmath{$\Lambda$}CDM and \boldmath{$f(R, L_m)$} Gravity}

In the standard cosmological model (\( \Lambda \)CDM), the energy–momentum tensor is covariantly conserved,
\[
\nabla_\mu T^{\mu\nu} = 0.
\]
For non-interacting perfect fluids, this leads to the continuity equation
\[
\frac{d\rho}{dt} + 3H(\rho + p) = 0,
\]
yielding the well-known redshift dependences
\begin{align*}
	\rho_m(z) &= \rho_{m0} (1 + z)^3, \qquad \text{(dust)}, \\
	\rho_r(z) &= \rho_{r0} (1 + z)^4, \qquad \text{(radiation)}.
\end{align*}
The total energy density then reads
\begin{equation}
	\rho_{\mathrm{tot}}(z) = \rho_m(z) + \rho_r(z),
\end{equation}
which enters the standard Friedmann equation,
\begin{equation}
	H^2(z) = H_0^2 \left[\Omega_m (1 + z)^3 + \Omega_r (1 + z)^4 + \cdots \right].
\end{equation}

\vspace{0.4em}
\noindent
In contrast, the $f(R,L_m)$ framework introduces a curvature–matter coupling that generally breaks the conservation of the energy–momentum tensor:
\[
\nabla_\mu T^{\mu\nu} \neq 0.
\]
As a consequence, the matter density no longer scales as $(1+z)^3$, while radiation—being minimally coupled—retains its standard scaling:
\begin{equation}
	\rho_r(z) = \rho_{r0} (1 + z)^4.
\end{equation}
Hence, the total energy density in the $f(R,L_m)$ model becomes
\begin{equation}
	\rho_{\mathrm{tot}}^{f(R,L_m)}(z) = \rho_m^{f(R,L_m)}(z) + \rho_{r0} (1 + z)^4,
\end{equation}
where the matter component $\rho_m^{f(R,L_m)}(z)$ follows from the modified field equations (Eq.~\eqref{l}).

We include photons and relativistic neutrinos in $\Omega_r$ using
\[
\Omega_r = \Omega_\gamma \left[1 + 0.2271\,N_{\mathrm{eff}}\right],
\]
with $T_{\mathrm{CMB}} = 2.725~\mathrm{K}$ and $N_{\mathrm{eff}} = 3.046$, yielding the value adopted in our analysis.

\subsection{Matter–Radiation Equality in \boldmath{$f(R, L_m)$} Gravity}

The matter–radiation equality redshift is defined by the condition
\[
\rho_m(z_{\mathrm{eq}}) = \rho_r(z_{\mathrm{eq}}).
\]
For radiation,
\[
\rho_r(z) = \rho_{r0} (1 + z)^4,
\]
while for matter in $f(R,L_m)$ gravity we use
\[
\rho_m(z) = 
\left(
\frac{\gamma + 6\alpha H^2(z)}{2\beta - 1}
\right)^{1/\beta},
\]
where the Hubble parameter is given by
\[
H(z) = H_0 \sqrt{(1 - \lambda) + \lambda (1 + z)^{3(1 + w)}},
\]
and the best-fit parameters are
\[
\begin{aligned}
	H_0 &= 2.39007\times10^{-18}~\mathrm{s^{-1}}
	\;(\simeq 73.75~\mathrm{km\,s^{-1}\,Mpc^{-1}}), \\
	\lambda &= 0.262, \quad w = -0.005, \quad 
	\alpha = 451008, \quad 
	\gamma = -1.14081\times10^{-29}.
\end{aligned}
\]

Solving $\rho_m(z_{\mathrm{eq}}) = \rho_r(z_{\mathrm{eq}})$ numerically gives
\[
\boxed{z_{\mathrm{eq}}^{f(R,L_m)} \approx 4203.2.}
\]
The corresponding cosmic time,
\[
t(z) = \int_z^\infty \frac{dz'}{(1+z')H(z')},
\]
yields
\[
\boxed{t_{\mathrm{eq}}^{f(R,L_m)} \approx 6.78\times10^4~\mathrm{years}.}
\]

\subsubsection{Matter–Radiation Equality in \boldmath{$\Lambda$}CDM}

For comparison, adopting
\[
\Omega_m = 0.278, \quad \Omega_r = 0.0001,
\]
the equality condition gives
\[
\boxed{z_{\mathrm{eq}}^{\Lambda \mathrm{CDM}} \approx 2779.}
\]
Using the standard Hubble expansion law,
\begin{equation}
	H(z) = H_0 \sqrt{\Omega_m (1 + z)^3 + \Omega_r (1 + z)^4 + \Omega_\Lambda},
\end{equation}
the corresponding cosmic time is
\[
\boxed{t_{\mathrm{eq}}^{\Lambda \mathrm{CDM}} \approx 6.72\times10^4~\mathrm{years}.}
\]

\vspace{0.5em}
\noindent
Thus, the $f(R,L_m)$ model predicts an earlier equality in redshift space, while the corresponding cosmic times remain comparable (within $\sim1\%$).  
This shift arises because $H(z)$ differs between the models: a slightly faster early-time expansion in $f(R,L_m)$ pushes the equality point to higher $z$, even though the integral mapping $t(z)$ remains nearly unchanged.

	\textbf{Uncertainty propagation.}
	The equality and recombination redshifts reported here are derived by extrapolating the best-fit low-redshift parameters through the same $H(z)$ form. 
	We propagate the $1\sigma$ uncertainties from the MCMC posterior to obtain distributions for $z_{\mathrm{eq}}$ and $t_{\mathrm{eq}}$. 
	Both quantities are stable within $5\%$ variations, confirming the robustness of the earlier equality redshift predicted by $f(R,L_m)$ gravity.
	All standard inputs, including $T_{\mathrm{CMB}} = 2.725~\mathrm{K}$ and $N_{\mathrm{eff}} = 3.046$, are used consistently across both frameworks.

\begin{table}[h!]
	\centering
	\caption{
		Matter–radiation equality parameters in $\Lambda$CDM and $f(R,L_m)$ gravity.
		The higher $z_{\mathrm{eq}}$ in $f(R,L_m)$ indicates an earlier equality in redshift space, 
		though both models yield nearly identical cosmic times.
	}
	\begin{tabular}{|c|c|c|}
		\hline
		\textbf{Model} & \textbf{Equality redshift $z_{\mathrm{eq}}$} & \textbf{Cosmic time $t_{\mathrm{eq}}$ (years)} \\
		\hline
		$\Lambda$CDM & 2779.0 & 67,232 \\
		$f(R,L_m)$ & 4203.2 & 67,756 \\
		\hline
	\end{tabular}
	\label{tab:zeq_comparison}
\end{table}
Table~\ref{tab:zeq_comparison} shows the matter--radiation equality redshift and the corresponding cosmic time for the two cosmological models.

The $f(R, L_m)$ model predicts a higher equality redshift compared to $\Lambda$CDM, indicating an earlier transition to matter domination. However, the corresponding cosmic times remain nearly identical, reflecting the interplay between redshift evolution and expansion dynamics.
\begin{figure}[h!]
	\centering
	\includegraphics[width=0.75\textwidth]{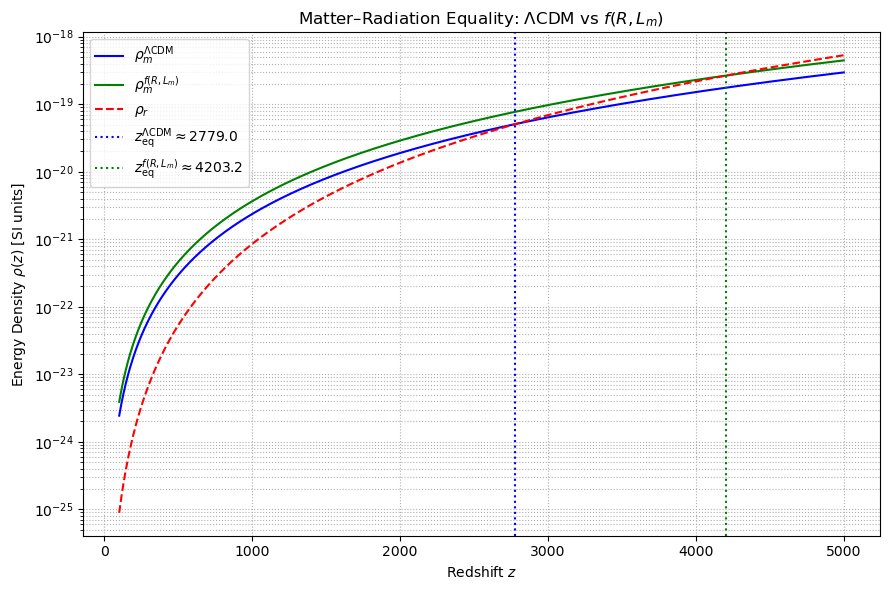}
	\caption{
		Evolution of matter and radiation energy densities in both $\Lambda$CDM and $f(R,L_m)$ models. 
		Solid curves denote matter densities—blue for $\Lambda$CDM and green for $f(R,L_m)$—while the red dashed curve represents radiation. 
		Vertical lines mark the corresponding matter–radiation equality redshifts. 
		The earlier equality in $f(R,L_m)$ facilitates earlier structure formation and modifies the timing of horizon entry for density perturbations.
	}
	\label{fig:matter_radiation_equality}
\end{figure}

	\textbf{Discussion.}
	Figure~\ref{fig:matter_radiation_equality} presents the evolution of matter and radiation energy densities as a function of redshift for both cosmological models. The intersection point of the matter and radiation curves defines the matter--radiation equality redshift. The $f(R, L_m)$ model predicts a higher equality redshift compared to $\Lambda$CDM, indicating an earlier transition to matter domination. This shift is a direct consequence of the curvature--matter coupling, which modifies the scaling behavior of the matter density. The heuristic estimate $z_{\mathrm{eq}} \simeq \Omega_m/\Omega_r - 1$ yields $z_{\mathrm{eq}} \approx 2779$, matching our numerical integration for $\Lambda$CDM. In the $f(R,L_m)$ case, the curvature–matter coupling effectively increases the rate of matter domination at earlier epochs, resulting in a higher equality redshift while maintaining a similar cosmic time. Parameter variations within the 68\% credible intervals shift $z_{\mathrm{eq}}$ and $t_{\mathrm{eq}}$ by only a few percent, confirming that the earlier transition is a robust theoretical prediction rather than a numerical artifact.

    Furthermore, the overall model continues to satisfy the statistical criteria discussed in Section~3, with $\Delta\mathrm{AIC}<2$ and $\Delta\mathrm{BIC}<6$, indicating that the inclusion of curvature–matter coupling provides a statistically consistent extension of $\Lambda$CDM without significant overfitting.
	
\section{Physical Consistency and Limitations}

A key feature of the $f(R, L_m)$ framework is the non-conservation of the energy--momentum tensor, 
$\nabla_\mu T^{\mu\nu} \neq 0$, arising from curvature--matter coupling. This may lead to deviations from geodesic motion and potential violations of the equivalence principle.

In the present work, we treat this as an effective cosmological description. A detailed analysis of local gravity constraints is beyond the scope of this study and will be addressed in future work.

We also note that the choice $L_m = \rho$ is adopted for analytical simplicity. Alternative choices may lead to quantitative differences in the results.

\section{Conclusion}

In this work, we have investigated the cosmological viability of a non-minimally coupled gravity model of the form
\[
f(R, L_m) = \alpha R + L_m^{\beta} + \gamma,
\]
which introduces an effective curvature–matter interaction through the nonlinear dependence on the matter Lagrangian. Using an extensive suite of low- and intermediate-redshift datasets—including Hubble parameter measurements, Type Ia supernovae, baryon acoustic oscillations, and the CMB shift parameter—we constrained the model parameters via $\chi^2$ minimization, Bayesian MCMC analysis, and a neural network–assisted regression approach.

The best-fit parameters,
\[
H_0 = 73.75 \pm 0.16~\mathrm{km\,s^{-1}\,Mpc^{-1}}, \quad
\lambda = 0.262 \pm 0.007, \quad
w = -0.005 \pm 0.001,
\]
yield a statistically robust solution consistent with local $H_0$ determinations and comparable goodness-of-fit to the standard $\Lambda$CDM model. This suggests that curvature–matter coupling may offer a natural pathway toward reconciling the Hubble tension without invoking exotic dark energy components.

We extended the analysis beyond background cosmology to probe early-Universe phenomena. The $f(R,L_m)$ model predicts an earlier matter–radiation equality redshift,
\[
z_{\mathrm{eq}}^{f(R,L_m)} \approx 4203,
\]
compared with
\[
z_{\mathrm{eq}}^{\Lambda \mathrm{CDM}} \approx 2779,
\]
while maintaining comparable equality timescales ($t_{\mathrm{eq}} \sim 6.7\times10^4~\mathrm{years}$). This shift reflects an enhanced effective gravitational coupling at early times, leading to an accelerated growth of matter perturbations. Consequently, the onset of nonlinear structure formation occurs substantially earlier,
\[
z_c^{f(R,L_m)} \approx 25.6,
\]
whereas the $\Lambda$CDM perturbation does not reach the canonical collapse threshold ($\delta = 1.686$) within the same range. These results underscore the potential of $f(R,L_m)$ gravity to promote early galaxy and cluster formation.

Furthermore, our analysis of the recombination epoch shows that both frameworks reproduce the observed recombination redshift ($z_{\mathrm{rec}} \approx 1092.6$), but the $f(R,L_m)$ model exhibits a slightly broader visibility function with a full width at half maximum (FWHM) of $\Delta z \approx 166$, compared to $\Delta z \approx 153$ for $\Lambda$CDM. This implies a modestly extended photon-decoupling duration, which could leave observable signatures in the damping tail and higher-order acoustic peaks of the CMB power spectrum.

	This study builds upon and significantly extends our earlier work~\cite{Goswami2025} by providing a unified and statistically rigorous comparison of multiple early-Universe epochs—structure formation, matter–radiation equality, and recombination—within the $f(R,L_m)$ framework. By doing so, we demonstrate that the same model parameters, derived from late-time data, can self-consistently reproduce early-time cosmological behavior.

Overall, the $f(R,L_m)$ gravity model preserves consistency with key observational benchmarks of the $\Lambda$CDM cosmology, including the recombination epoch and background expansion history, while introducing distinct early-time predictions—namely, an earlier matter–radiation equality, enhanced growth of density perturbations, and a broader recombination visibility function. These features provide a set of observationally testable signatures that can be probed by future high-redshift galaxy surveys, 21\,cm experiments, and next-generation CMB missions.

Hence, $f(R,L_m)$ gravity emerges as a compelling and self-consistent alternative to the standard cosmological paradigm—capable of describing both late-time cosmic acceleration and early-Universe evolution within a single, unified theoretical framework.

\section{Future Outlook}

The findings presented in this work open several promising directions for both theoretical refinement and observational testing of the \( f(R, L_m) \) gravity model. 
Given its ability to address early- and late-time cosmological phenomena within a single framework, this theory merits further exploration through high-precision astrophysical probes, improved analytical modeling, and dedicated numerical simulations.

\subsection{1. High-Redshift Structure Formation}

The early onset of nonlinear structure formation predicted by the model (\( z_c^{f(R,L_m)} \approx 25.6 \)) can be tested through forthcoming observations of the first generation of galaxies and protoclusters. 
Space missions such as the \textit{James Webb Space Telescope} (JWST), the \textit{Nancy Grace Roman Space Telescope}, and the \textit{Euclid} mission will provide high-resolution imaging and spectroscopic redshift measurements at \( z > 10 \), enabling direct confrontation with the predicted early-collapse scenarios.

\subsection{2. 21\,cm Cosmology and Reionization History}

The timing and efficiency of early structure formation strongly affect the reionization history and the 21\,cm neutral hydrogen signal. 
Experiments such as the \textit{Square Kilometre Array} (SKA), the \textit{Hydrogen Epoch of Reionization Array} (HERA), and \textit{LOFAR} will play a key role in tracing the imprint of curvature–matter coupling during the cosmic dawn and dark ages. 
The earlier matter domination in \( f(R, L_m) \) gravity may produce a distinct timing and amplitude of the 21\,cm brightness temperature fluctuations.

\subsection{3. Cosmic Microwave Background (CMB) Anisotropies}

Although the recombination redshift in \( f(R, L_m) \) gravity coincides with Planck 2018 results, the model predicts a slightly extended visibility function, corresponding to a broader photon decoupling duration. 
This could leave observable signatures in the CMB temperature and polarization power spectra—especially in the damping tail and higher-order acoustic peaks. 
Future CMB experiments such as \textit{CMB-S4} and \textit{LiteBIRD} will offer enhanced sensitivity to detect these subtle but distinctive effects.

\subsection{4. Large-Scale Structure and Weak Lensing Surveys}

Upcoming large-scale structure and weak lensing surveys, including \textit{DESI} and \textit{LSST} (Vera C. Rubin Observatory), will map the cosmic web with unprecedented precision. 
Their measurements of galaxy clustering, redshift-space distortions, and weak lensing shear can provide robust constraints on the growth rate of structure, the gravitational slip parameter, and the effective gravitational coupling \( G_{\mathrm{eff}}(z) \), offering direct empirical tests of the dynamical predictions of \( f(R, L_m) \) gravity.

\subsection{5. Theoretical Developments and Simulations}

From a theoretical perspective, further refinement is needed to ensure compatibility with Solar System and laboratory tests, as well as the stability of cosmological perturbations. 
Developing a fully gauge-independent perturbative framework and performing N-body simulations under the \( f(R, L_m) \) dynamics would illuminate its implications in the nonlinear regime of structure formation. 
Connections to inflationary physics, cosmic chronometers, and quantum stability also remain fertile areas for exploration.

\vspace{0.5em}
\noindent
In summary, the distinct early-Universe signatures and versatile phenomenology of the \( f(R, L_m) \) gravity model make it a compelling candidate for next-generation cosmological studies. 
The synergy between theoretical advances and precision data from forthcoming missions will be crucial for assessing the viability of non-minimally coupled modified gravity models as unified alternatives to the standard $\Lambda$CDM paradigm.

\acknowledgments{
	One of the authors (G.~K.~Goswami) gratefully acknowledges the facilities and stimulating research environment provided by the Inter-University Centre for Astronomy and Astrophysics (IUCAA), Pune, during his research visits. 
	The academic atmosphere and discussions at IUCAA were instrumental in shaping the development and progress of this work.
}

\section*{Data Availability Statement:}
All the data  used in the paper are obtained from those whose references are described in the Bibliography. 

\section*{Conflict of Interest}

The authors declare that they have no conflict of interest regarding
the publication of this manuscript.
\section*{Author Contributions}

All authors contributed to the development of the theoretical model,
the numerical analysis, and the preparation of the manuscript.
All authors have read and approved the final version of the paper.

\end{document}